\documentclass[10pt]{article}
\usepackage{latexsym,graphicx}
\newcommand{\be}{\begin{equation}}
\newcommand{\ee}{\end{equation}}
\def\n{\noindent}
\catcode `\@=11
\catcode `\@=12
\begin{document}
\begin{center}
\large{\bf {NONSINGULAR COSMOLOGICAL MODELS WITH A VARIABLE COSMOLOGICAL TERM $\Lambda$ }} \\
\vspace{10mm}
\normalsize{ANIRUDH PRADHAN $^{a,c,}$\footnote{Corresponding Author}, KASHIKA SRIVASTAVA$^{a,d}$ and AMRIT LAL AHUJA$^{b}$} \\
\vspace{2.5mm}
\normalsize{$^{a}$\it{Department of Mathematics, Hindu Post-graduate College, 
 Zamania-232 331, Ghazipur, India}} \\
\vspace{2.5mm}
\normalsize{$^{b}$\it{ Inter-University Centre for Astronomy and Astrophysics, Ganeshkhind,
 Pune-411 007, India}}\\
\vspace{2.5mm}
{\it {E-Addresses: $^{c}$pradhan@iucaa.ernet.in ;  $^{b}$ahuja@iucaa.ernet.in;  
$^{d}$kash\_gg@yahoo.co.uk}}\\
\end{center}
\vspace{10mm}
\begin{abstract} 
Exact solutions of the Einstein's field equations describing a spherically 
symmetric cosmological model without a big bang or any other kind of singularity 
recently obtained by Dadhich and Patel (2000) are revisited. The matter content 
of the model is a shear-free perfect fluid with isotropic pressure and a radial 
heat flux. Three different exact solutions are obtained for both perfect fluid and 
fluid with bulk viscosity. It turns out that the cosmological rerm $\Lambda(t)$ is a 
decreasing function of time, which is consistent with recent observations of type 
Ia supernovae. 
\end{abstract}
\smallskip
\n PACS number: 98.80.Es, 98.80.-k\\
\n Keywords : non-singular models, cosmology, variable cosmological constant, causality 
principle\\
\newpage
\section{Introduction}
\noindent
The problem of cosmological singularity is one of the most fundamental issues
in modern theoretical cosmology. Due to the powerful singularity 
theorem \cite{ref1,ref2}, it was widely believed that cosmological models must have  
initial singularity. However, in 1990 Senovilla \cite{ref3} obtained the first 
singularity-free cosmological perfect-fluid (with a realistic equation of state 
$3p = \rho$) solution of the Einstein equation and since then the possibility of 
constructing regular cosmologies was renewed. The interest for regular cosmologies
had stifled for nearly $30$ years due to the powerful singularity theorems, which
seemed to preclude such spacetimes under very general requirements, such as
chronology protecting, energy and generic conditions. The open way to regular
cosmologies was found in the violation of some technical premises of the theorems.
The remarkable feature is the absence of an initial singularity, the curvature and 
matter invariants being regular and smooth everywhere. This corresponds to a 
cylindrically-symmetric spacetime filled with an isotropic radiation perfect fluid. 
For instance, it was shown by Chinea {\it et al.}\cite{ref4} that the Senovilla 
spacetime did not possess a compact achronal set without edge and could not have 
closed trapped surfaces. However, the first results were not encouraging. The 
extension of the Senovilla solution to a family of spacetime left the set of 
regular models limited to a zero-measure subset surrounded by spacetime with Ricci 
and Weyl curvature singularities \cite{ref5}. A thorough discussion of the model 
such type  can be found in Senovilla \cite{ref6}. This family is shown to be included in a 
wider class of separable cosmological models, which comprises FLRW universe \cite{ref7}. 
Other properties of these solutions, such as their inflationary behaviour, generalized 
Hubble law and the feasibility of constructing a realistic non-singular cosmological model 
are studied therein. 
\newline
\par
A large family of non-singular cosmological models and generalization thereof have 
been considered but they all are cylindrically symmetric \cite{ref8} $-$ \cite{ref10}. 
For practical cosmology the spherical symmetry, however, is more appropriate. It 
is therefore pertinent to seek spherically symmetric nonsingular models. The first 
model of this kind was obtained by Dadhich \cite{ref11} with an imperfect fluid 
with a heat flux. The model satisfied all energy conditions and had no singularity 
of any kind. Dadhich {\it et al.}\cite{ref12} also obtained a non-singular model 
with null radiation flux. These models are both inhomogeneous and anisotropic and 
have a typical behaviour beginning with two density at $t \rightarrow - \infty$, 
contracting to high density at $t = 0$ and then again expanding to low density at 
$t \rightarrow  \infty$. An interesting feature of the spacetime metric of these 
models is that it contains an arbitrary function of time which can be constrained 
to comply with the demand of non-singularity and energy conditions, Dadhich and 
Raychaudhuri \cite{ref13} later showed how a particle choice of this function leads 
to a model of an ever existing spherically symmetric universe, oscillating between 
two regular states, which involves blue shifts as in the quasi steady state 
cosmological model of Hoyle, Burbige and Narlikar \cite{ref14} and is filled with 
a non-adiabatic fluid with anisotropic pressure and radial heat flux. These 
observations led to the search of spherically symmetric singularity-free 
cosmological models with a perfect fluid source characterized by isotropic pressure. 
Due to this search Tikekar \cite{ref15} constructed two spherically symmetric 
singularity-free relativistic cosmological models, describing universes filled 
with non-adiabatic perfect fluid accompanied by heat flow along radial direction.
Recently, many researchers \cite{ref16} $-$ \cite{ref20} have studied non-singular 
cosmological models in different context. From a purely theoretical point of 
view, the investigation of nonsingular cosmological models gives invaluable insight
into the spacetime structure, the inherent nonlinear character of gravity and its 
interaction with matter fields. As a by-product it also deepens our understanding
of the singularity theorem, in particular the assumptions lying in their base \cite{ref7}.
\newline
\par
Models with a dynamic cosmological term $\Lambda(t)$ are becoming popular as 
they solve the cosmological-constant problem in a natural way. There are significant 
observational evidence for the detection of Einstein's cosmological constant, 
$\Lambda$ or a component of material content of the universe, that varies slowly 
with time and space and so acts like $\Lambda$. New observations of type Ia supernovae 
(Garnavich {\it et al.} \cite{ref21}, Perlmutter {\it et al.} \cite{ref22}, Riess 
{\it et al.} \cite{ref23}, Schmidt {\it et al.} \cite{ref24}), cosmic microwave 
background (CMB) anisotropies (e.g., Lineweaver \cite{ref25}), galaxy surveys (e.g., Lahav 
and Bridle \cite{ref26}), gravitational lensing (e.g., Chiba and Yoshii \cite{ref27}), and the 
Ly$\alpha$ forest (e.g., Weinberg {\it et al.} \cite{ref28}) argue for a nonzero 
cosmological ``constant'' with $\Omega_{\Lambda} (\equiv \Lambda/3H^{2}_{0})
\approx 0.6-0.7$. This quantity may {\it not be constant} as has been 
appreciated for $30$ years (Bergmann \cite{ref29}, Wagoner \cite{ref30}, 
Linde \cite{ref31}, Kazanas \cite{ref32}). Scalar fields (Dolgov \cite{ref33}, 
Abbott \cite{ref34}, Barr \cite{ref35}, Peeble and Ratra \cite{ref36}, 
Friemann {\it et al.} \cite{ref37}, Moffat \cite{ref38}, Starobinsky \cite{ref39}), 
tensor fields (Hawking \cite{ref40}, Dolgov \cite{ref41}), D-branes (Ellis, 
Mavromatos and Nanopoulos \cite{ref42}), nonlocal effects (Banks \cite{ref43}, 
Linde \cite{ref44}), wormholes (Coleman \cite{ref45}), inflationary mechanisms 
(Brandenberger and Zhitnitsky \cite{ref46}, Peebles and Vilenkin \cite{ref47}), 
and cosmological perturbations (Abramo, Brandenberger and Mukhanov \cite{ref48}) 
have all been shown to give rise an effective cosmological term that decays 
with time. Earlier researches on this topic, are contained in Zeldovich \cite{ref49}, 
Weinberg \cite{ref50}, Dolgov \cite{ref51}, Bertolami \cite{ref52}, Felten and 
Isaacman \cite{ref53}, Charlton and Turner \cite{ref54}, Sandaga \cite{ref55}, 
Carroll, Press and Turner \cite{ref56}. Some of the recent discussions on the 
cosmological-constant ``problem'' and consequences on cosmology with a time-varying 
cosmological-constant have been discussed by Dolgov and Silk \cite{ref57}, Sahni and 
Starobinsky \cite{ref58}, Peebles \cite{ref59}, Padmanabhan \cite{ref60}, Carroll \cite{ref61}, 
Vishwakarma \cite{ref62}, and Pradhan {\it et al.}\cite{ref63}. This 
motivates us to study the cosmological models, where $\Lambda$ varies with time.  
\newline
\par
Recently Dadhich and Patel \cite{ref64} obtained a shear-free nonsingular spherical 
model with heat flux. This model satisfies the weak and strong energy conditions and 
also has a physically acceptable fall-off behaviour in both $r$ and $t$ for physical 
and kinematic parameters. In this paper, motivated by the situation discussed above, 
we shall focus on the problem with varying cosmological constant in presence of 
perfect fluid and also in presence of bulk viscous fluid. We do this by extending
the work of Dadhich and Patel \cite{ref64} by including varying cosmological constant.
The remainder of this paper is organized as follows. In Section $2$ we give a description 
of the cosmological models with its dynamical equations and solve them under the initial
conditions inspired by Dadhich and Patel. We also investigate three different cosmological
models to different values of the function $P(t)$ and discuss results for these regimes.
Section $3$ comprises bulk viscous universe. We present our discussions and conclusions 
in Section $4$.  

\section{A Perfect Fluid Universe Revisited}
\label{sec:pfur}
In this section, we review the solutions obtained by Dadhich and Patel \cite{ref64}.
The metric of the model is given in the form
\begin{equation} 
\label{eq1}  
ds^{2} = (r^{2} + P)^{2n} dt^{2} - (r^{2} + P)^{2m} \left[dr^{2} + r^{2}(d\theta^{2}
+ \sin^{2}\theta d\phi^{2})\right],
\end{equation}
where \\
\[
2n = 2m \pm \sqrt{8m^{2} + 8m + 1}, \\
\]
in particular,
\begin{equation} 
\label{eq2}
2m = 1 - \sqrt{\frac{3}{2}} < 0, ~ ~ 2n = \sqrt{\frac{3}{2}}.   
\end{equation}
Here $P = P(t)$ which can be chosen freely. The Einstein field equations for a perfect 
fluid with time-dependent cosmological constant and a redial heat flux read:
\begin{equation} 
\label{eq3}
R_{ik} - \frac{1}{2} R g_{ik} + \Lambda g_{ik} = - \left[(\rho + p)u_{i}u_{k} - p g_{ik}
+ \frac{1}{2}(q_{i}u_{k} + q_{k}u_{i})\right],
\end{equation}
where we have set $\frac{8\pi G}{c^{2}} = 1$, $u_{i}u^{i} = 1 = -q_{i}q^{i}$, 
$q_{i}u^{i} = 0$, $\rho$ and $p$ denote the fluid density and isotropic pressure, 
and $q_{i}$ is the radial heat flux vector. \\
From Equations (\ref{eq1}) and (\ref{eq3}) we obtain
\begin{equation} 
\label{eq4}
\rho = \frac{3m^{2}\dot{P}^{2}}{(r^{2} + P)^{2n + 2}} - \frac{4m\{3P + (m +1)r^{2}\}}
{(r^{2} + P)^{2m + 2}} + \Lambda,
\end{equation}
\[
p = - \frac{m}{(r^{2} + P)^{2n + 2}}\left[2(r^{2} + P)\ddot{P} + (3m - 2n -2)\dot{P}^
{2}\right]
\]
\begin{equation} 
\label{eq5}
+ \frac{4}{(r^{2} + P)^{2m + 2}}\left[(m + n)P + n^{2}r^{2}\right] - \Lambda,
\end{equation}
\begin{equation} 
\label{eq6}
q = \frac{4m(n + 1)r \dot{P}}{(r^{2} + P)^{n + 2}},
\end{equation}
where $q_{i} = q g^{1}_{i}$. The expansion and acceleration are obtained as
\begin{equation} 
\label{eq7}
\theta = \frac{3m\dot{P}}{(r^{2} + P)^{n + 1}},
\end{equation}
\begin{equation} 
\label{eq8}
\dot{u}_{r} = - \frac{nr}{r^{2} + P}.
\end{equation}
We have freedom of choosing the function $P(t)$ so that to give a non-singular 
behaviour to the above parameters. As a matter of fact, there are multiple choices
(see, Dadhich and Patel \cite{ref64}), for instant, $P(t) = a^{2} + b^{2}t^{2}$, 
$a^{2} + e^{-bt^{2}}$, $a^{2} + b^{2}\cos \omega t$, $a^{2} > b^{2}$. For all these
choices it is observed that all the physical and kinematic parameters remain regular and
finite for the entire range of variables. Note that the model admits an interesting 
oscillating behaviour in time, with oscillations between two finite regular states. 
Oscillating nonsingular models are quite novel and interesting in their own accord. \\
For complete determinacy of the system, we assume an equation of state of the form
\begin{equation} 
\label{eq9}
p = \gamma \rho, ~ ~ 0 \leq \gamma \leq 1,
\end{equation}
where $\gamma$ is a constant. \\

\subsection{Model 1:}
\label{sec:mdl1}
We set $P(t) = a^{2} + b^{2} t^{2}, ~ ~ a^{2} > b^{2}$.
In this case the matter density $\rho$, the fluid pressure $p$, the heat flux 
parameter $q$ and kinematic parameter of expansion $\theta$ are found to have 
the following expressions:
\begin{equation} 
\label{eq10}
\rho = \frac{12m^{2}b^{4}t^{2}}{(r^{2} + a^{2} + b^{2}t^{2})^{2n + 2}} -
\frac{4m\left[3(a^{2} + b^{2}t^{2}) + (m + 1)r^{2}\right]}{(r^{2} + a^{2} 
+ b^{2}t^{2})^{2m + 2}} + \Lambda,
\end{equation}
\begin{equation} 
\label{eq11}
p = - \frac{4mb^{2}\left[r^{2} + a^{2} + (3m - 2n -1)b^{2}t^{2}\right]}{(r^{2} + a^{2} 
+ b^{2}t^{2})^{2n + 2}} + \frac{4\left[(m + n)(a^{2} + b^{2}t^{2}) + n^{2}r^{2}\right]}
{(r^{2} + a^{2} + b^{2}t^{2})^{2m + 2}} - \Lambda,
\end{equation}
\begin{equation} 
\label{eq12}
q = \frac{8m(n + 1)rb^{2}t}{(r^{2} + a^{2} + b^{2}t^{2})^{n + 2}},
\end{equation}
\begin{equation} 
\label{eq13}
\theta = \frac{6mb^{2}t}{(r^{2} + a^{2} + b^{2}t^{2})^{n + 1}}.
\end{equation}
Equations (\ref{eq10}) and (\ref{eq11}), with the use of (\ref{eq9}), reduce to 
\[
(1 + \gamma)\Lambda = - \frac{4mb^{2}\left[r^{2} + a^{2} + (3m - 2n - 1 + 3 m 
\gamma)b^{2} t^{2}\right]}{(r^{2} + a^{2} + b^{2}t^{2})^{2n + 2}}
\]
\begin{equation} 
\label{eq14}
+ \frac{4\left[(m + n + 3m\gamma)(a^{2} + b^{2}t^{2}) + \{n^{2} + m(m + 1)
\gamma \}r^{2}\right]}{(r^{2} + a^{2} + b^{2}t^{2})^{2m + 2}}.
\end{equation}
\begin{figure}
\begin{center}
\includegraphics[angle=-90, width=0.8\textwidth]{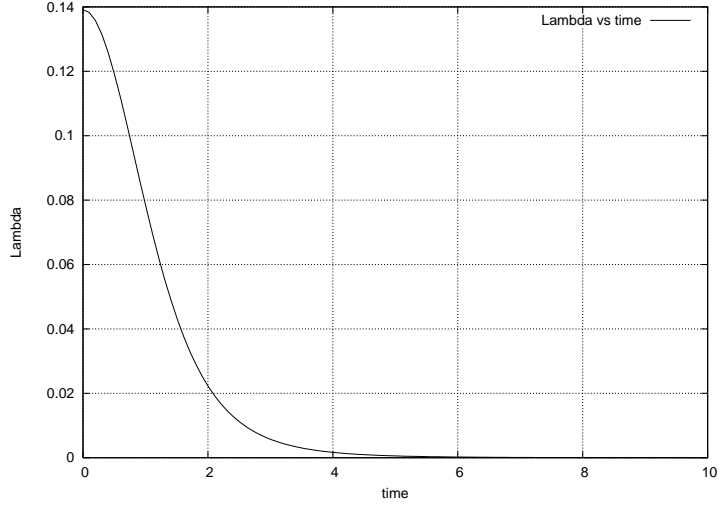}
\caption[Model 1]{Variation of $\Lambda$ with time for 2.1 Model 1. The values of parameters 
are: $\rm m=1, ~n=1+\sqrt{17}/2, ~\gamma =0.5, ~a=2, ~b=1 ~and ~r=1$.}
\label{fig:plt1}
\end{center}
\end{figure}
From above equations it is evident that the matter density is always and 
everywhere positive while positivity of pressure is ensured, if $3m -2n < 1$. 
The heat flux parameter $q > 0$ for $t < 0$, $q = 0$ for $t = 0$ and $q < 0$
for $t > 0$. Equation (\ref{eq13}) implies that the model describes an expanding 
universe for $t < 0$ with $q > 0$ and a contracting universe for $t > 0$ for
$q < 0$, the switching from contracting phase to phase of expansion occurring 
at $t = 0$. 

From Equation (\ref{eq14}) we observe that the cosmological constant 
$\Lambda$ is a decreasing function of time (see Figure \ref{fig:plt1}). 
We also observe that the $\Lambda$ approaches a small and positive value 
at late times which is supported by recent type Ia supernova observations \cite{ref21} $-$ \cite{ref24}.

\subsection{Model 2:}
\label{sec:mdl2}
We set $P(t) = a^{2} + e^{-bt^{2}},~ ~ a^{2} > b^{2}$.
In this case the matter density $\rho$, the fluid pressure $p$, the heat flux 
parameter $q$ and kinematic parameter of expansion $\theta$ are found to have 
the following expressions:
\begin{equation} 
\label{eq15}
\rho = \frac{12m^{2}b^{2}t^{2}e^{-2bt^{2}}}{(r^{2} + a^{2} + e^{-bt^{2}})^{2n + 2}} -
\frac{4m \left[3(a^{2} + e^{-bt^{2}}) + (m + 1)r^{2}\right]}{(r^{2} + a^{2} + 
e^{-bt^{2}})^{2m + 2}} + \Lambda,
\end{equation}
\[
p = - \frac{4mbe^{-bt^{2}}\left[(r^{2} + a^{2})(2bt^{2} - 1) - e^{-bt^{2}} + 
(3m - 2n)bt^{2}e^{-bt^{2}}\right]}{(r^{2} + a^{2} + e^{-bt^{2}})^{2n + 2}}
\]
\begin{equation} 
\label{eq16}
+ \frac{4\left[(m + n)(a^{2} + e^{-bt^{2}}) + n^{2}r^{2}\right]}{(r^{2} + a^{2} + 
e^{-bt^{2}})^{2m + 2}} - \Lambda,
\end{equation}
\begin{equation} 
\label{eq17}
q = - \frac{8mbr(n + 1)te^{-bt^{2}}}{(r^{2} + a^{2} + e^{-bt^{2}})^{n + 2}},
\end{equation}
\begin{equation} 
\label{eq18}
\theta = - \frac{6mbte^{-bt^{2}}}{(r^{2} + a^{2} + e^{-bt^{2}})^{n + 1}}.
\end{equation}
By using Equation (\ref{eq9}) and eliminating $\rho(t)$ between (\ref{eq15}) and (\ref{eq16}), 
we obtain
\[
(1 + \gamma)\Lambda = - \frac{4mbe^{-bt^{2}}\left[\{3(1 + \gamma)m - 2n\}bt^{2}e^{-bt^{2}} 
+ (r^{2} + a^{2})(2bt^{2} - 1) - e^{-bt^{2}}\right]}{(r^{2} + a^{2} + e^{-bt^{2}})^{2n + 2}}
\]
\begin{equation} 
\label{eq19} 
+ \frac{4\left[\{(1 + 3\gamma)m + n\}(a^{2} + e^{-bt^{2}}) + \{m(m + 1)\gamma + n^{2}\}r^{2}\right]}
{(r^{2} + a^{2} + e^{-bt^{2}})^{2m + 2}}.
\end{equation}
\begin{figure}
\begin{center}
\includegraphics[angle=-90, width=0.8\textwidth]{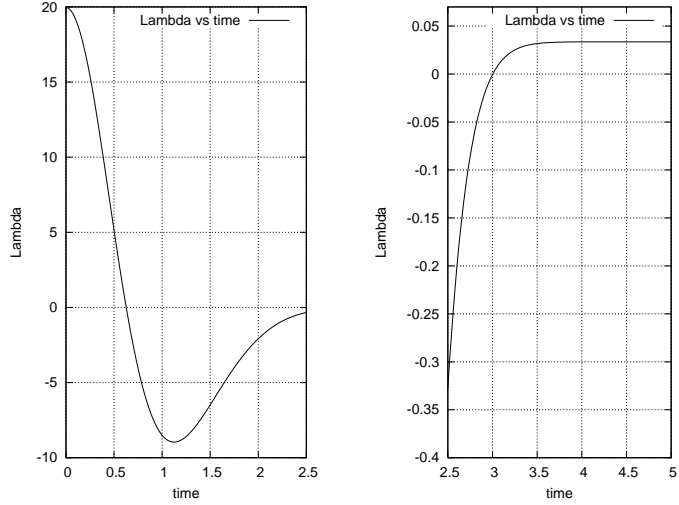}
\caption[Model 2]{Variation of $\Lambda$ with time for 2.2 Model 2. The parameters are: $\rm m=1, 
~n=1-\sqrt{17}/2, ~\gamma =0.5, ~r=1, ~a=2 ~and ~b=1$.}
\label{fig:plt2}
\end{center}
\end{figure}
From above equations it is observed that the matter density is always and everywhere
positive while positivity of pressure is ensured if $3m - 2n > 0$. Thus it is also observed
that the requirements of weak and strong energy conditions are fulfilled throughout
the spacetime of this model. The dominant energy condition, which requires $\rho \geq p$,
cannot, however, be satisfied: it is clearly violated for large $r$. Thus this model 
satisfies weak and strong but not the dominant energy condition. 

From Equation ({\ref{eq19}) , it is observed that the $\Lambda$ 
first decreases, reaches to a negative value then increases and 
becomes a constant small positive value (see Figure \ref{fig:plt2}). 
This could play the role of dark energy.

\subsection{Model 3:}
\label{sec:mdl3}
We set $P(t) = a^{2} + b^{2}\cos \omega t, ~ ~ a^{2} > b^{2}$.
In this case the matter density $\rho$, the fluid pressure $p$, the heat flux 
parameter $q$ and kinematic parameter of expansion $\theta$ are found to have 
the following expressions:
\begin{equation} 
\label{eq20}
\rho = \frac{3m^{2}b^{4}\omega^{2} \sin^{2}\omega t}{(r^{2} + a^{2} + b^{2}\cos \omega t)
^{2n + 2}} - \frac{4m[3(a^{2} + b^{2}\cos \omega t) + (m + 1)r^{2}]}{(r^{2} + a^{2} + b^{2}
\cos \omega t)^{2m + 2}} + \Lambda,
\end{equation}
\[
p = \frac{mb^{2}\omega^{2}\left[2(r^{2} + a^{2} + b^{2}\cos \omega t) \cos \omega t - 
(3m - 2n -2)\sin^{2} \omega t \right] } {(r^{2} + a^{2} + b^{2}\cos \omega t)^{2n + 2}}
\]
\begin{equation} 
\label{eq21}
+ \frac{4[(m + n)(a^{2} + b^{2}\cos \omega t) + n^{2}r^{2}]}{(r^{2} + a^{2} + b^{2}
\cos \omega t)^{2m + 2}} - \Lambda,
\end{equation}
\begin{equation} 
\label{eq22}
q = - \frac{4m(n + 1) r b^{2} \omega \sin \omega t}{(r^{2} + a^{2} + b^{2}
\cos \omega t)^{n + 2}},
\end{equation}
\begin{equation} 
\label{eq23}
\theta = - \frac{3 b^{2} \omega \sin \omega t}{(r^{2} + a^{2} + b^{2}\cos \omega t)^{n + 1}}.
\end{equation}
Equations (\ref{eq20}) and (\ref{eq21}), with the use of (\ref{eq9}, give 
\[
(1 + \gamma)\Lambda = - \frac{mb^{2}\omega^{2}\left[\{3(1 + b^{2}\gamma)m - 2\left(n + 1\right)\}\sin^{2}
\omega t - 2(r^{2} + a^{2} + b^{2}\cos \omega t)\cos\omega t \right]}{(r^{2} + a^{2} + 
b^{2}\cos \omega t)^{2n + 2}}
\]
\begin{equation} 
\label{eq24} 
+ \frac{4\left[\{(1 + 3\gamma)m + n\}(a^{2} + b^{2}\cos\omega t) + \{m(m + 1)\gamma  
+ n^{2}\}r^{2}\right]}{(r^{2} + a^{2} + b^{2}\cos\omega t)^{2m + 2}}.
\end{equation}

\begin{figure}
\begin{center}
\includegraphics[angle=-90, width=0.8\textwidth]{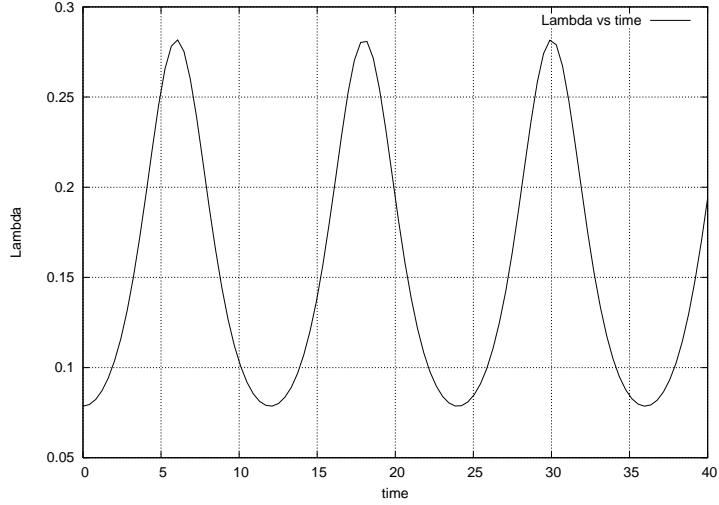}
\caption[Model 3]{Variation of $\Lambda$ with time for 2.3 Model 3. Here $\rm m=1, 
~n=1+\sqrt{17}/2, ~\gamma =0.5, ~r=1, ~a=2, ~b=1 ~and ~\omega = \pi /6$.}
\label{fig:plt3}
\end{center}
\end{figure}
From the above equations it is observed that the matter density is always positive whereas
the pressure is non-negative if $3m - 2n < 0$. Thus it can be seen that the requirements
of weak and strong energy conditions are fulfilled throughout the spacetime of this model 
but not the dominant energy condition. From equation (\ref{eq23}) the expansion parameter 
indicates that the universe of this model in the phase of contraction for $2 \alpha \pi < 
\omega t < (2 \alpha + 1)\pi$ where $\alpha$ takes on integer values only. During the phase
of contracting $q < 0$ while during the expansion phase $q > 0$ while $q$ vanishing when 
switching from contraction to expansion occurs vice-versa. 

From equation ({\ref{eq24}), it is observed that the $\Lambda$ is 
oscillating due to properties of sinusoidal functions 
(see Figure \ref{fig:plt3}). It is also worth-noting that the average value 
(with respect to one period) of $\Lambda$ is positive. Here we will have 
negative equation of state at late times required to support the current 
acceleration of universe.

\section {Bulk Viscous Universe}
The equations of bulk viscosity can be obtained from the general relativistic
field equation, we replace the effective pressure\cite{ref49}
\begin{equation} 
\label{eq25}
\bar{p} = p - \xi \theta, 
\end{equation}
where $p$ is the pressure due to perfect fluid present, $\xi$ is the coefficient of
bulk viscosity and $\theta$ is the expansion scalar. Thus, given $\xi(t)$ we can
solve for cosmological parameters. In most of the investigations, involving
bulk viscosity, it is assumed to be a simple power function of the energy density
(Pavon \cite{ref65}, Maartens \cite{ref66}, Zimdahl \cite{ref67})
\begin{equation}
\label{eq26}
\xi(t) = \xi_{0} \rho^{k},
\end{equation}
where $\xi_{0}$ and $k$ are constants. If $k = 1$, equation (\ref{eq26}) may correspond
to a radiative fluid (Weinberg\cite{ref50}). However, more realistic models 
(Santos\cite{ref68}) are based on $k$ lying in range $0 \leq k\leq \frac{1}{2}$. \\
Introducing (\ref{eq25}) and (\ref{eq26}) into (\ref{eq5}), we obtain
\[
p = \frac{3m\xi_{0}\rho^{k}\dot{P}}{(r^{2} + P)^{n + 1}} - \frac{m}{(r^{2} + P)^{2n + 2}}
\left[2(r^{2} + P)\ddot{P} + (3m - 2n -2)\dot{P}^{2}\right]
\]
\begin{equation} 
\label{eq27}
+ \frac{4}{(r^{2} + P)^{2m + 2}}\left[(m + n)P + n^{2}r^{2}\right] - \Lambda.
\end{equation}

\subsection {Model 1:} 
We set $P(t) = a^{2} + b^{2} t^{2}, ~ ~ a^{2} > b^{2}$. In this case we consider two
following cases. 

\subsubsection {Case I : solution for $\xi = \xi_{0}$}
When $k = 0$, Equation (\ref{eq26}) reduces to $\xi = \xi_{0}$ (constant) and hence Equation 
(\ref{eq27}) with the help (\ref{eq9}) and (\ref{eq10}) reduces to the form
\[
(1 + \gamma)\rho = \frac{6mb^{2}\xi_{0} t}{(r^{2} + a^{2} + b^{2}t^{2})^{n + 1}} +
\frac{4mb^{2}\left[(2n + 1)b^{2}t^{2} - (r^{2} + a^{2})\right]}{(r^{2} + a^{2} + 
b^{2}t^{2})^{2n + 2}}
\]
\begin{equation} 
\label{eq28}
- \frac{4\left[(2m - n)(a^{2} + b^{2}t^{2}) + \{m(m + 1) - n^{2}\}r^{2}\right]}
{(r^{2} + a^{2} + b^{2}t^{2})^{2m + 2}}.
\end{equation}
Eliminating $\rho(t)$ between Equations (\ref{eq10}) and (\ref{eq28}), we get 
\[
(1 + \gamma)\Lambda = \frac{6mb^{2}\xi_{0} t}{(r^{2} + a^{2} + b^{2}t^{2})^{n + 1}} - 
\frac{4mb^{2}\left[r^{2} + a^{2} + \{3(1 + \gamma)m - 2n - 1\}b^{2} t^{2}\right]}{(r^{2} 
+ a^{2} + b^{2}t^{2})^{2n + 2}}
\]
\begin{equation} 
\label{eq29}
+ \frac{4\left[\{(1 + 3\gamma)m + n\}(a^{2} + b^{2}t^{2}) + \{n^{2} + m(m + 1)
\gamma \}r^{2}\right]}{(r^{2} + a^{2} + b^{2}t^{2})^{2m + 2}}.
\end{equation}

\subsubsection {Case II : solution for $\xi = \xi_{0} \rho$}
When $k = 1$, Equation (\ref{eq26}) reduces to $\xi = \xi_{0} \rho $ and hence Equation 
(\ref{eq27}) with the help of Equations (\ref{eq9}) and (\ref{eq10}) takes the form
\[
\left[1 + \gamma - \frac{6\xi_{0} m b^{2} t}{(r^{2} + a^{2} + b^{2}t^{2})^{n + 1}}
\right]\rho = \frac{4mb^{2}\{(2n + 1)b^{2}t^{2} - (r^{2} + a^{2})\}}{(r^{2} + a^{2} + 
b^{2}t^{2})^{2n + 2}}
\]
\begin{equation} 
\label{eq30}
+ \frac{4\left[(n - 2m)(a^{2} + b^{2}t^{2}) + r^{2}\{n^{2} - m(m + 1)\}\right]}
{(r^{2} + a^{2} + b^{2}t^{2})^{2m + 2}}.
\end{equation}
Eliminating $\rho(t)$ between Equations (\ref{eq30}) and (\ref{eq10}), we get
\[
[(1 + \gamma)(r^{2} + a^{2} + b^{2}t^{2})^{n + 1} - 6\xi_{0} mb^{2}t]\Lambda = 
\]
\[
24\xi_{0}m^{2}b^{2}t \times \left[\frac{3mb^{4}t^{2}}{(r^{2} + a^{2} + b^{2}t^{2})^{2n + 2}} 
- \frac{3(a^{2} + b^{2}t^{2}) + (m + 1)r^{2}}{(r^{2} + a^{2} + b^{2}t^{2})^{2m + 2}}\right]
\]
\[
- \frac{4mb^{2}\left[r^{2} + a^{2} + \{3(1 + \gamma)m - 2n -1\}b^{2}t^{2}\right]}{(r^{2} + 
a^{2} + b^{2}t^{2})^{n + 1}} + 
\]
\begin{equation} 
\label{eq31}
\frac{4\left[\{(1 + 3\gamma)m + n\}(a^{2} + b^{2}t^{2}) + 
(n^{2} - m^{2} + 1)r^{2}\right]}{(r^{2} + a^{2} + b^{2}t^{2})^{2m - n + 1}}. 
\end{equation}
\begin{figure}
\begin{center}
\includegraphics[angle=-90, width=0.8\textwidth]{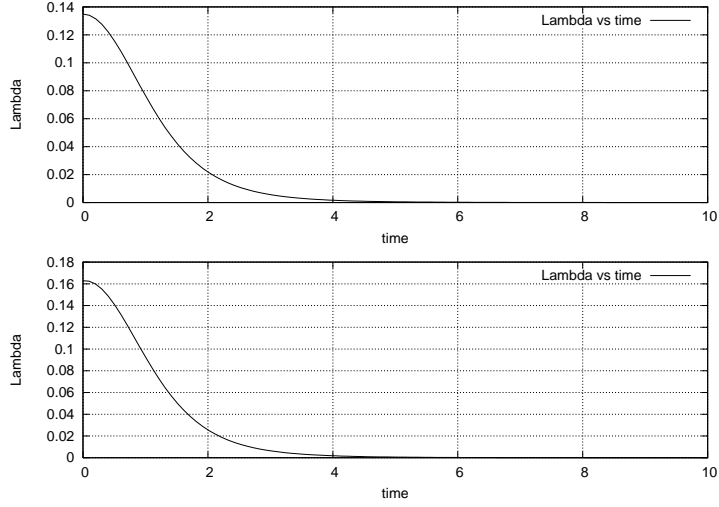}
\caption[Model 3.1]{Variation of $\Lambda$ with time for 3.1 Model 1, case I (lower panel) 
and II (upper panel). Here $\rm m=1, ~n=1+\sqrt{17}/2, ~\gamma =0.5, ~r=1, ~a=2, ~b=1 ~and 
~\xi_0 = 1$.}
\label{fig:plt4}
\end{center}
\end{figure}
From Equations (\ref{eq29}) and (\ref{eq31}), we observe that the 
$\Lambda$ is a decreasing function of time (see Figure \ref{fig:plt4}), 
and it approaches a small positive value which is similar to the previously
discussed Model 1 (Section \ref{sec:mdl1}). 

\subsection{Model 2:}
We set $P(t) = a^{2} + e^{-bt^{2}}, ~ ~ a^{2} > b^{2}$. In this case we consider two 
following cases.

\subsubsection {Case I : solution for $\xi = \xi_{0}$}
When $k = 0$, Equation (\ref{eq26}) reduces to $\xi = \xi_{0}$(constant) and hence Equation 
(\ref{eq27}) with the help of (\ref{eq9}) and (\ref{eq15}) reduces to the form
\[
(1 + \gamma)\rho = -\frac{6mb\xi_{0}te^{-bt^{2}}}{(r^{2} + a^{2} + e^{-bt^{2}})^{n + 1}} + 
\]
\[
\frac{4mbe^{-bt^{2}}\left[(2nbt^{2} + 1)e^{-bt^{2}} - (r^{2} + a^{2})(2bt^{2} - 1)\right]}
{(r^{2} + a^{2} + e^{-bt^{2}})^{2n + 2}}
\]
\begin{equation} 
\label{eq32}
- \frac{4\left[(2m - n)(a^{2} + e^{-bt^{2}}) + \{m(m + 1) - n^{2}\}r^{2}\right]}{(r^{2} + a^{2} + 
e^{-bt^{2}})^{2m + 2}}.
\end{equation}
Eliminating $\rho(t)$ between Equations (\ref{eq15}) and (\ref{eq32}), we get
\[
(1 + \gamma)\Lambda =  - \frac{6mb\xi_{0}te^{-bt^{2}}}{(r^{2} + a^{2} + e^{-bt^{2}})
^{n + 1}} 
\]
\[
- \frac{4mbe^{-bt^{2}}\left[\{3(1 + \gamma)m - 2n\}bt^{2}e^{-bt^{2}} 
+ (r^{2} + a^{2})(2bt^{2} - 1) - e^{-bt^{2}}\right]}{(r^{2} + a^{2} + e^{-bt^{2}})
^{2n + 2}}
\]
\begin{equation} 
\label{eq33} 
+ \frac{4\left[\{(1 + 3\gamma)m + n\}(a^{2} + e^{-bt^{2}}) + \{m(m + 1)\gamma + n^{2}\}r^{2}
\right]}{(r^{2} + a^{2} + e^{-bt^{2}})^{2m + 2}}.
\end{equation}

\subsubsection {Case II : solution for $\xi = \xi_{0} \rho$}
When $k = 1$, Equation (\ref{eq26}) reduces to $\xi = \xi_{0}\rho$ and hence Equation 
(\ref{eq27}) takes the form
\[
\left[\gamma - \frac{6mb\xi_{0}e^{-bt^{2}}}{(r^{2} + a^{2} + e^{-bt^{2}})^{n + 1}}\right]
\rho =  
\]
\[
- \frac{4mbe^{-bt^{2}}\left[(r^{2} + a^{2})(2bt^{2} - 1) - e^{-bt^{2}} + 
(3m - 2n)bt^{2}e^{-bt^{2}}\right]}{(r^{2} + a^{2} + e^{-bt^{2}})^{2n + 2}}
\]
\begin{equation} 
\label{eq34}
+ \frac{4\left[(m + n)(a^{2} + e^{-bt^{2}}) + n^{2}r^{2}\right]}{(r^{2} + a^{2} + 
e^{-bt^{2}})^{2m + 2}}.
\end{equation}
Eliminating $\rho(t)$ between Equations (\ref{eq34}) and (\ref{eq15}), we get
\[
[(1 + \gamma)(r^{2} + a^{2} + e^{-bt^{2}})^{n + 1} - 6mb\xi_{0}e^{-bt^{2}}]\Lambda =
\]
\[
24m^{2}b\xi_{0}e^{-bt^{2}}\left[\frac{3mb^{2}t^{2}e^{-2bt^{2}}}{(r^{2} + a^{2} + 
e^{-bt^{2}})^{2n + 2}} - \frac{3(a^{2} + e^{-bt^{2}}) + (m + 1)r^{2}}{(r^{2} + a^{2} 
+ e^{-bt^{2}})^{2m + 2}}\right]
\]
\[
- \frac{4mbe^{-bt^{2}}\left[\{3(1 + \gamma)m - 2n \}bt^{2}e^{-bt^{2}} + (r^{2} + a^{2})
(2bt^{2} - 1) - e^{-bt^{2}}\right]}{(r^{2} + a^{2} + e^{-bt^{2}})^{n + 1}}
\]
\begin{equation} 
\label{eq35}
 + \frac{4\left[\{(1 + 3\gamma)m + n\}(a^{2} + e^{-bt^{2}}) + \{m(m + 1)\gamma + 
n^{2}\}r^{2}\right]}{(r^{2} + a^{2} + e^{-bt^{2}})^{2m - n + 1}}.
\end{equation}
\begin{figure}
\begin{center}
\includegraphics[angle=-90, width=0.8\textwidth]{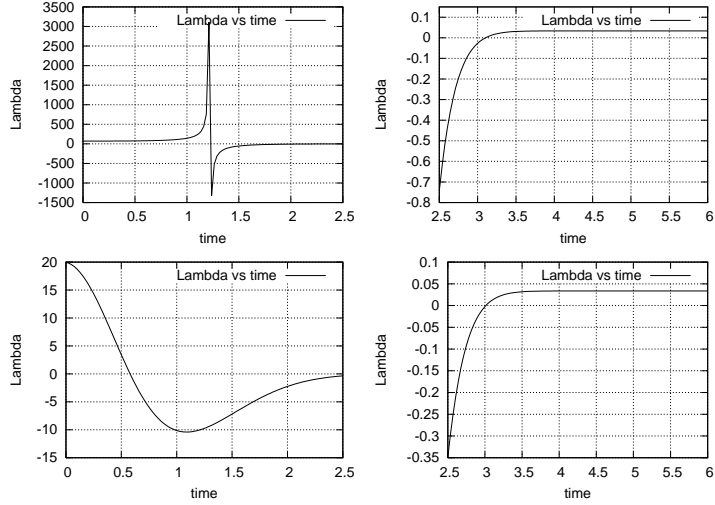}
\caption[Model 3.2]{Variation of $\Lambda$ with time for 3.2 Model 2, cases I (lower panel)
and II (upper panel). Here $\rm m=1, ~n=1-\sqrt{17}/2, ~\gamma =0.5, ~r=1, ~a=2, ~b=1 ~and 
~\xi_0 = 1$.}
\label{fig:plt5}
\end{center}
\end{figure}
From Equation (\ref{eq33}) we observe that the value of $\Lambda$ first increases 
slowly and suddenly reaches to peak, then it has sharp decrease to a negative value, 
again, it has a slow increment and finally becomes a small positive constant value 
(see Figure \ref{fig:plt5} upper panel). From Equation (\ref{eq35}) we observe that 
the $\Lambda$ first decreases and then increases and finally approaches to a small 
positive constant. This could play the role of dark energy.
\subsection{Model 3:}
We set $P(t) = a^{2} + b^{2} \cos \omega t,  ~ ~ a^{2} > b^{2}$. In this case we 
consider two following cases.

\subsubsection {Case I : solution for $\xi = \xi_{0}$}
When $k = 0$, Equation (\ref{eq26}) reduces to $\xi = \xi_{0}$(constant) and hence Equation
(\ref{eq27}) with the help of (\ref{eq9}) and (\ref{eq20}) reduces to  the form
\[
(1 + \gamma)\rho  = \frac{3\omega b^{2}\xi_{0} \sin \omega t}{(r^{2} + a^{2} + b^{2}\cos \omega t)
^{n + 1}} + 
\]
\[
\frac{mb^{2}\omega^{2}\left[\{3m(b^{2} - 1) + 2(n + 1)\}\sin^{2} \omega t + 2(r^{2} + a^{2} 
+ b^{2}\cos \omega t)\cos \omega t \right] } {(r^{2} + a^{2} + b^{2}\cos \omega t)^{2n + 2}}
\]
\begin{equation} 
\label{eq36}
- \frac{4[(a^{2} + b^{2}\cos \omega t)(2m - n) - n^{2}r^{2}]}{(r^{2} + a^{2} + b^{2}
\cos \omega t)^{2m + 2}}.
\end{equation}
Eliminating $\rho(t)$ between Equations (\ref{eq20}) and (\ref{eq36}), we get
\[
(1 + \gamma)\Lambda = \frac{3\omega b^{2}\xi_{0} \sin \omega t}{(r^{2} + a^{2} + 
b^{2}\cos \omega t)^{n + 1}} 
\]
\[
- \frac{mb^{2}\omega^{2}\left[\{3(1 + b^{2}\gamma)m - 2n\}\sin^{2}\omega t - 2(r^{2} 
+ a^{2} + b^{2}\cos \omega t)\cos\omega t \right]}{(r^{2} + a^{2} + b^{2}\cos \omega t)^{2n + 2}}
\]
\begin{equation} 
\label{eq37} 
+ \frac{4\left[\{(1 + 3\gamma)m + n\}(a^{2} + b^{2}\cos\omega t) + \{m(m + 1)\gamma  
+ n^{2}\}r^{2}\right]}{(r^{2} + a^{2} + b^{2}\cos\omega t)^{2m + 2}}.
\end{equation}
\begin{figure}
\begin{center}
\includegraphics[angle=-90, width=0.8\textwidth]{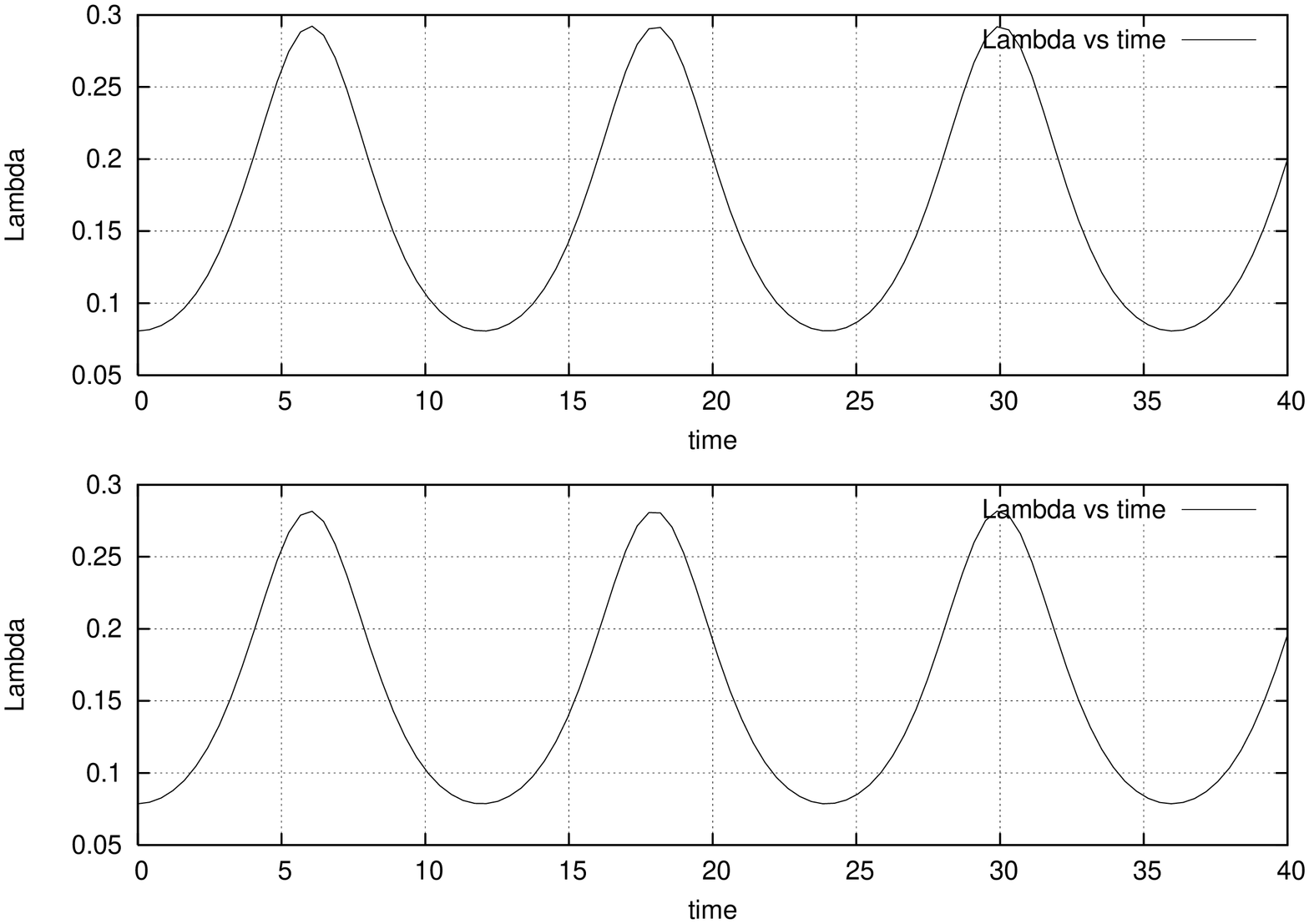}
\caption[Model 3.3]{Variation of $\Lambda$ with time for 3.3 Model 3, case I (lower panel) and
II (upper panel). Here $\rm m=1,
 ~n=1+\sqrt{17}/2, ~\gamma =0.5, ~r=1, ~a=2, ~b=1, ~\xi_0 = 1, ~\omega =\pi/6$.}
\label{fig:plt6}
\end{center}
\end{figure}

\subsubsection {Case II : solution for $\xi = \xi_{0} \rho$}
When $k = 1$, Equation (\ref{eq26}) reduces to $\xi = \xi_{0}\rho$ and hence Equation 
(\ref{eq27}) takes the form
\[
\left[\gamma - \frac{3\xi_{0}b^{2}\omega \sin \omega t} {(r^{2} + a^{2} + b^{2}
\cos \omega t)^{n + 1}}\right]\rho =  
\]
\[
\frac{mb^{2}\omega^{2}\left[2(r^{2} + a^{2} + b^{2}\cos \omega t) \cos \omega t - 
(3m - 2n -2)\sin^{2} \omega t \right] } {(r^{2} + a^{2} + b^{2}\cos \omega t)^{2n + 2}}
\]
\begin{equation} 
\label{eq38}
+ \frac{4[(m + n)(a^{2} + b^{2}\cos \omega t) + n^{2}r^{2}]}{(r^{2} + a^{2} + b^{2}
\cos \omega t)^{2m + 2}}.
\end{equation}
Eliminating $\rho(t)$ between Equations (\ref{eq37}) and (\ref{eq20}), we obtain
\[
\left[(1 + \gamma)(r^{2} + a^{2} + b^{2}\cos \omega t)^{n + 1} + 3\xi_{0}b^{2} \omega 
\sin \omega t\right] \Lambda = 
\]
\[
+ 3 \xi_{0}b^{2}m\omega \sin \omega t\left[\frac{3mb^{4}\omega^{2}\sin^{2}\omega t}
{(r^{2} + a^{2} + b^{2}\cos \omega t)^{2n + 2}} - \frac{4\{3(a^{2} + b^{2}\cos \omega t)
+ (m + 1)r^{2}\}}{(r^{2} + a^{2} + b^{2}\cos \omega t)^{2m + 2}}\right]
\]
\[
+ \frac{mb^{2}\omega^{2}\left[2(r^{2} + a^{2} + b^{2}\cos \omega t)\cos \omega t -
\{3m(1 - \gamma b^{2}) - 2(n + 1)\}\sin^{2}\omega t\right]} {(r^{2} + a^{2} + b^{2}
\cos \omega t)^{n + 1}}
\]
\begin{equation}
\label{eq39}
+ \frac{4\left[\{(1 + 3\gamma)m + n\}(a^{2} + b^{2}\cos \omega t) + 
(m + n^{2} + 1)r^{2}\right]}{(r^{2} + a^{2} + b^{2}\cos \omega t)^{2m - n + 1}}.
\end{equation}
From Equations (\ref{eq37})  and (\ref{eq39}) we observe that the $\Lambda$ oscillates 
with time due to properties of sinusoidal functions, present in these equations. The nature 
of these models are same as already discussed in Model 3 (Section \ref{sec:mdl3}).

\section {Conclusions}
We have obtained a new class of spherically symmetric inhomogeneous cosmological 
models with a perfect fluid and also a bulk viscous fluid as the source of matter 
with a radial heat flux without a big bang or any other singularity. These are the 
shear-free nonsingular models. These are inhomogeneous and hence accelerating but 
not shearing. It is the heat flux that combines with pressure gradient to avoid 
singularity. From the point of view of realistic cosmology, these models share with 
the standard FRW model, spherical symmetry and the absence of shear. \\
\newline
\par
The cosmological constant is a parameter describing the energy density of the vacuum
(empty space), and a potentially important contribution to the dynamical history
of the universe. The physical interpretation of the cosmological constant as 
vacuum energy is supported by the existence of the ''zero point'' energy predicted 
by quantum mechanics. In quantum mechanics, particle and antiparticle pairs are 
consistently being created out of the vacuum. Even though these particles exist 
for only a short amount of time before annihilating each other they do give the 
vacuum a non-zero potential energy. In general relativity, all forms of energy 
should gravitate, including the energy of vacuum, hence the cosmological constant.
A negative cosmological constant adds to the attractive gravity of matter, therefore
universes with a negative cosmological constant are invariably doomed to re-collapse \cite{ref69}.
A positive cosmological constant resists the attractive gravity of matter
due to its negative pressure. For most universes, the positive cosmological constant
eventually dominates over the attraction of matter and drives the universe to expand 
exponentially \cite{ref70}.
\newline
\par
The cosmological constants in all models given in Sections $2.1$ and $3.1$ are decreasing
functions of time and they all approach a small and positive value at late times
which are supported by the results from recent type Ia supernova 
observations recently obtained by the High-z Supernova Team and Supernova Cosmological
Project (Garnavich {\it et al.} \cite{ref21}, Perlmutter {\it et al.} \cite{ref22},
Riess {\it et al.} \cite{ref23}, Schmidt {\it et al.} \cite{ref24}). Thus, with our approach,
we obtain a physically relevant decay law for the cosmological term unlike other
investigators where {\it adhoc} laws were used to arrive at a mathematical expressions 
for the decaying vacuum energy. Our derived models provide a good agreement with the
observational results. We have derived value for the cosmological constant $\Lambda$
and attempted to formulate a physical interpretation for it.\\
\newline
\par
This paper adds a novel family of shear-free models to Senovilla's first model \cite{ref3} 
and a large family of cylindrical nonsingular models \cite{ref5,ref7,ref8,ref10}
and a large family of spherical nonsingular models \cite{ref11,ref12,ref64} avoiding cosmic 
singularity in the absence of shear.    
\section*{Acknowledgements} 
 The authors wish to thank the Inter-University Center for Astronomy and
Astrophysics, Pune, India, for warm hospitality and excellent facilities where this 
work was done. We also thank to Mohammed Sami for his fruitful suggestions and comments.\\
\newline

\end{document}